\documentstyle[12pt]{article}
\def\beqa{\begin{eqnarray}}
\def\eeqa{\end{eqnarray}}
\def\beq{\begin{equation}}
\def\eeq{\end{equation}}
\begin{document}
\def\bib#1{[{\ref{#1}}]}
\begin{titlepage}
     \title{Thin shell quantization in Weyl spacetime}

      \author{{S. Capozziello$^a$\thanks{E-mail: capozziello@sa.infn.it},
A. Feoli$^{a,b}$\thanks{E-mail: feoli@unisannio.it},  G.
Lambiase$^{a}$\thanks{E-mail: lambiase@sa.infn.it}, G.
Papini$^{c}$\thanks{E-mail: papini@uregina.ca}}
\\ {\em {\small $^a$Dipartimento di Scienze Fisiche "E. R. Caianiello",}}
\\ {\em {\small Universit\`a di Salerno, I--84081 Baronissi (Sa), Italy.}}
\\ {\em {\small INFN, Sezione di Napoli, Gruppo Collegato di Salerno,
Italy.}}
\\ {\em {\small $^b$Facolt\`a di Ingegneria, Universit\`a del Sannio.}}
\\ {\em {\small $^c$Department of Physics, University of Regina,}}
\\ {\em {\small Regina, Sask. S4S 0A2, Canada.}}}
          \date{\today}
          \maketitle

              \begin{abstract}
We study the problem of quantization of thin shells in a
Weyl--Dirac theory by deriving a Wheeler--DeWitt equation from the
dynamics. Solutions are found which have interpretations in both
cosmology and particle physics.

          \end{abstract}

\vspace{20. mm} PACS: 98.80.Hw, 04.50.+h, 4.20.Cv

\vfill

\end{titlepage}

\section{\normalsize \bf Introduction}
Phase transitions and spontaneous symmetry breaking play a leading
role in physics and account for a large number of phenomena. In
cosmology they suggest improvements to the standard cosmological
model, while in particle physics, they give faithful accounts of
the behaviour and relationship of the fundamental interactions.

In cosmology, in particular, the mechanism which fosters the
transition from false to true vacuum can account for the change
from a de Sitter to a Friedmann--Robertson--Walker universe and
can be enacted in a variety of ways.   We can have first--order
phase transitions \cite{kolb} or higher--order transitions
\cite{linde}, but the aim always is to realize a transition from a
totally symmetric spacetime, where the matter content is due to
quantum fields in a polarized vacuum state, to our asymmetric
world where we observe four fundamental interactions and three
families of particles.

The Glashow--Weinberg--Salam model \cite{kaku} and the Guth's
inflation \cite{guth} are the starting points of a plethora of
models whose number is then restricted by the accuarcy of sky and
ground--based observations.

Symmetry breaking also is relevant in quantum gravity where  it
can give insight into topological changes and  is useful in
dealing with quantum gravitational fluctuations. Such fluctuations
may induce a minimum length, for example, thus introducing an
additional source of uncertainty in physics. Furthermore, if the
geometry is subject to quantum fluctuations, these give rise to a
spacetime foam at the Planck scale \cite{garay}. By understanding
the quantum evolution of this foam and by the definition of its
Hamiltonian structure, the problem of quantum gravity could become
more tractable, if not completely solvable.

The breaking of some geometrical symmetry can be considered also
classically from a dynamical point of view. In \cite{wood}, the
breaking of conformal invariance has been used in order to
construct bubbles in a Weyl spacetime.

These can be associated with microscopic particles \cite{wood1},
but  can be as well considered in a cosmological context. Their
classical dynamics is discussed in \cite{feoli}.We study the issue
of their quantization in this work.

The problem of the quantization of (false-) vacuum bubbles,
  using the formalism of
thin shells in general gelativity, was first studied
  by Berezin et al.\cite{berezin} and then
  by several authors \cite{berezin1}.
The different results obtained depend upon
 the different ways used to construct  the Hamiltonian
  structure. Recently, Zloshchastiev \cite{zloshchastiev}
 has shown that quantizing a conservation
 law  is equivalent to introducing a Hamiltonian in a minisuperspace in the
spirit
 of the Wheeler--De Witt
quantization. Accordingly, it is possible to quantize thin
 shells directly from the equations of motion, if the time does not appear
 explicitly, thus avoiding the problems of time--slicing
and time gauge.

In this note,   we  apply the approach of \cite{zloshchastiev} to
the Weyl--Dirac model discussed  in \cite{wood},\cite{feoli}. In
Sect.2, we  outline the main features of the classical model.
Sect.3 is devoted to the quantization of the model along the lines
of the Wheeler--De Witt approach. Several solutions are given. In
Sect.4, the results are discussed.

\section{\normalsize\bf The classical model.}

 Bubbles with an interior de Sitter geometry can be constructed
dynamically by means of the Gauss--Mainardi--Codazzi (GMC)
formalism, assuming a conformally--invariant exterior geometry.
Specifically, one can use \cite{dirac}
 a Weyl--Dirac theory  of the form
 \begin{equation}
\label{1} I_{D}=\int \left[-\frac{1}{4}f_{\mu \nu }f^{\mu \nu }+
\beta ^{2}R+6\beta _{,\mu }\beta ^{,\mu }- \lambda \beta
^{4}\right]\sqrt{-g}d^{4}x\,.
\end{equation}
 Here $f_{\mu \nu }=\kappa _{\nu ,\mu }-\kappa _{\mu ,\nu }$ is
the electromagnetic field and $\beta$ is a nonminimally coupled
scalar field. The equations of motion derived from action
(\ref{1}) are
\begin{equation}
\Box_{\nu}f^{\mu\nu}=0\,,
\end{equation}
which are the usual Maxwell equations, and
\begin{equation}
R_{\mu\nu}-\frac{1}{2}g_{\mu\nu}R=\frac{1}{2\beta^2}E_{\mu\nu}+I_{\mu\nu}+
\frac{1}{2}\lambda g_{\mu\nu}\beta^2\equiv T_{\mu\nu}\,,
\end{equation}
which represent the Einstein equations where, $E_{\mu\nu}$ is the
Maxwell tensor, and
\begin{equation}
I_{\mu\nu}=\frac{2}{\beta}(\Box_{\nu}\Box_{\mu}\beta-
g_{\mu\nu}\Box_{\alpha}\Box^{\alpha}\beta)-\frac{1}{\beta^2}
(4\beta_{,\mu}\beta_{,\nu}-g_{\mu\nu}\beta_{,\alpha}\beta^{,\alpha})\,,
\end{equation}
is the stress--energy tensor of the scalar field. $\Box_{\nu}$ is
the Riemannian covariant derivative. The units are  such that
$c^3(16\pi G)^{-1}=1$.

As shown below, the scalar field $\beta$ provides the surface
tension of the bubbles.

By breaking
  conformal symmetry in a spherical region of Weyl spacetime, it is
 possible to construct a stable bubble of standard Riemannian geometry
 (Minkowski, de Sitter or anti--de Sitter spacetimes) that can
represent either an
 elementary particle or an entire universe , depending on the scale of the parameters.
 The most general spherically symmetric line element is given
by
\begin{equation}
ds^{2}=-e^{\nu (r,t)}dt^{2}+e^{\mu (r,t)}dr^{2}+ r^{2}(d\theta
^{2}+\sin ^{2}\theta d\phi ^{2}).
\end{equation}
The exterior and interior geometries are distinguished by writing
$t_{E,I}$, $\nu _{E,I}$ and $\mu _{E,I}$ in the exterior and
interior spacetimes $V^{E,I}$, respectively.
  The GMC formalism can be used
to connect the two different spacetime
 regions separated by a hypersurface
 $\Sigma$.  In this formalism, one introduces the
spherically symmetric intrinsic metric
\begin{equation}
ds_{\Sigma }^{2}=-d\tau ^{2}+R^{2}(\tau )(d\theta ^{2}+ \sin
^{2}\theta d\phi ^{2})
\end{equation}
and imposes suitable conditions on the scalar field $\beta$
 in order to obtain the equation of motion of the thin shell.
  Actually, the condition
$\beta = \beta _{0}$ in $V^{I}$, where $\beta_{0}$ is a constant,
together with the boundary condition that $\beta $ is continuous
across $\Sigma $ for all $t$, require that $\beta $ be a constant
with respect to the intrinsic time $\tau $ of the thin shell
defined at $r = R(\tau )$. That is,
\begin{equation}
\frac{d\beta }{d\tau }|_{r=R} = \beta _{,t}X_{E} + \beta ' \dot{R}
= 0,
\end{equation}
where $X_{E}\equiv dt/d\tau $ and the prime and dot denote
differentiation with respect to $r$ and $\tau $, respectively.
  All the mathematical details
 of the model can be found in \cite{wood},\cite{wood1},\cite{feoli}.

As the Birkhoff theorem is violated in Weyl spacetime, it also is
possible to obtain  a spherically symmetric solution depending on
time \cite{feoli}. This situation is extremely interesting in
cosmology since $R(t)$ can assume the role of the scale factor in
a bubbly universe which is undergoing a phase transition after a
symmetry breaking.

Solving the field equations, one finds the exact solutions for the
interior and external geometries. One obtains,  for the interior
metric,
\begin{equation}
e^{-\mu _{I}}=1+\frac{1}{6}\lambda \beta _{0}^{2}r^{2}= e^{\nu
_{I}},
\end{equation}
while in Weyl spacetime, one gets
\begin{equation} e^{\nu _{E}}=-\frac{\beta
_{,t}^{2}}{\beta '^{2}(1+ \frac{1}{6}\lambda \beta ^{2}r^{2})}
\end{equation}
and
\begin{equation}
e^{\mu _{E}} = -\frac{1+\frac{1}{6}\lambda \beta ^{2}r^{2}}{\gamma
^{2} \beta ^{2}r^{4}},
\end{equation}
where $\gamma $ is an arbitrary constant with dimension
(length)$^{-1}$. If one assumes that the sign of the constant
$\lambda $ does not change during the formation of the bubble,
then the sign must be taken to be negative to ensure the correct
signature of the metric (see \cite{wood} and \cite{feoli} for the
conventions used here) so the Riemannian geometry is a de Sitter
spacetime. Since the electromagnetic field vanishes in $V^I$, the
interior stress--energy tensor reduces to
\begin{equation}
\label{matter} T^I_{\mu\nu}=\frac{1}{2}\lambda
g_{\mu\nu}\beta^2_0\,,
\end{equation}
and one recovers an effective cosmological constant.

These solutions can then be used to determine the equation of
motion for $r = R(\tau )$
 in the frame comoving with the thin shell where the proper time is $\tau$.
 In this case, the equation of motion acquires the simple form
\cite{feoli}
\begin{equation}
\label{motion} \dot R^2 = {\alpha^2 \left(
1-R^2/R^2_{eq}\right)\over \left( R^2_{eq}/ R^2-1\right)^2
-\alpha^2}\,,
\end{equation}
where $\alpha^{2} = \gamma ^{2}\beta _{0}^{2}R_{eq}^{4}$. From
$\dot{R}=0$, one obtains the condition
\begin{equation}
\label{equivalenza} R_{eq} = \frac{1}{\beta
_{0}}\sqrt{\frac{6}{|\lambda |}}\,.
\end{equation}
It follows, from Eq.({\ref{equivalenza}}), that the size of the
shell is governed by the value of the scalar field inside it and
by the parameter $\lambda$. In other words, it is the effective
cosmological constant
\begin{equation}
\Lambda=\frac{1}{2}\lambda\beta_0^{2}\,,
\end{equation}
which rules the size of the bubble, as it must be in any
first--order inflationary model \cite{kolb}.

Eq.(\ref{motion}) can be solved in several interesting cases
\cite{feoli}.
 Bubbles which are created in
the Weyl vacuum with initial radii near either of the endpoints of
the interval $0<R<R_{eq}/\sqrt{2}$ exist indefinitely with a
finite radius or else collapse to $R=0$ while bubbles that are
created in the Weyl vacuum, with an initial radius greater than
$R_{eq}$, appear to be unstable in all cases, except for the
trivial case $R=R_{eq}$.
\section{\normalsize\bf The quantum model}
Turnig now to the problem of  quantization of the above dynamical
model,  we restrict ourselves to a minisuperspace approach,
following \cite{zloshchastiev}, where an immediate quantization of
conservation laws is given.

We can start from the pointlike Lagrangian
\begin{equation}
\label{lagrange} {\cal
L}=\frac{1}{2}m_{pl}\dot{R}^2-\frac{m_{pl}}{2}
 \left[
{\alpha^{2}(R^{2}/R^{2}_{eq}-1)\over\left( R^2_{eq}/ R^2 -
1\right)^2 -\alpha^2}\right]\,,
\end{equation}
which is defined on the tangent space ${\cal
TQ}\equiv\{R,\dot{R}\}$, which is just a one--dimensional
minisuperspace (the only variable is $R$). Here $m_{pl}$ is the
Planck mass. Eq.(\ref{motion}) is easily obtained as a first order
integral of dynamics. The classical solutions are those in
\cite{feoli}. Let us now define the canonical momentum
\begin{equation}
\label{momentum} \dot R^2 = {\pi^2\over m^2_{pl}}\,.
\end{equation}
Eqs.(\ref{lagrange}) and (\ref{momentum}) yield the Hamiltonian
\begin{equation}
\label{hamilton} {\cal H}=\frac{\pi^2}{2m_{pl}}
 + \frac{m_{pl}}{2} \left[
{\alpha^{2}(R^{2}/R^{2}_{eq}-1)\over\left( R^2_{eq}/ R^2 -
1\right)^2 -\alpha^2}\right]\,.
\end{equation}
We can then set the energy function related to the Lagrangian
equal to zero, i.e.
\begin{equation}
E_{\cal L}=\frac{\partial {\cal L}}{\partial\dot{R}}\dot{R}-{\cal
L}=0\,.
\end{equation}
This implies that ${\cal H}=0$ on the trajectories  (\ref{motion})
and  ${\cal H}$ can then be interpreted as the Hamiltonian
constraint of the Wheeler--De Witt approach. By a canonical
quantization of the momentum (\ref{momentum}), i.e.
$\pi=-i\partial_{R}$, we get the Wheeler--De Witt equation
\begin{equation}
\label{wdw} H|\Psi\rangle = 0\qquad \Longrightarrow\qquad
\left\{\frac{\partial^2}{\partial R^2} + m^2_{pl} \left[
{\alpha^{2}(1-R^{2}/R^{2}_{eq})\over\left( R^2_{eq}/ R^2 -
1\right)^2 -\alpha^2}\right]\right\} |\Psi\rangle=0\,.
\end{equation}
By setting
\begin{equation}
 x= {R\over R_{eq}}\,,\qquad a=m_{pl}R_{eq}\alpha\,,
\end{equation}
 Eq.(\ref{wdw}) becomes
\begin{equation}
\label{wdw1}
 \Psi'' + \left[{a^2(1-x^{2}) x^{4}\over
(1-x^{2})^{2} - \alpha^{2}x^{4}}\right] \Psi =0\,,
\end{equation}
where primes indicate differentiation with respect to $R$, and the
ket--vector $|\Psi\rangle$ has been replaced with the functional
$\Psi$ for simplicity.

At this point, an important remark is necessary. The quantization of thin
shells can be obtained from several effective Lagrangians so that
the corresponding quantum theory is not uniquely defined. This fact
implies that the Wheeler--De Witt equation can be achieved in many
different ways due, for example,  to the choice of factor ordering
and momentum operator). In our case, following
\cite{zloshchastiev},\cite{vilenkin},\cite{carugno}, we have done
the simplest choices. Furthermore,  we are going to take into account
the semiclassical limit so that a more complicate approach is not
necessary for the following discussion of solutions.

The solutions of Eq.(\ref{wdw1}) give the probability amplitude to
get bubbles of a given size, mass and energy. The term inside the
square brakets is the superpotential of the model which defines
the classical/quantum boundary of the minisuperspace.   It
separates, in principle,  the Euclidean from the Lorentzian zones.
In the first case,  the wavefunction is {\it under} the
superpotential barrier and has an exponential behaviour. In the
latter case, it has an oscillating behaviour and is over or above
the barrier as in the case of standard quantum tunneling. The
separation of these zones is given by the zeros of the
superpotential. In our case, they are $R=\pm  R_{eq}$ and $R=0$.

Approximate solutions can be found from the analysis of
Eq.(\ref{wdw1}). The asymptotic behaviours are particularly
interesting.

 For $|x|<<1$, we have
\begin{equation}
\label{wdw2} \Psi'' + a^2 x^{4}\Psi = 0\,,
\end{equation}
whose solution is a superposition of Bessel functions $Z_{\nu}(z)$
given by
\begin{equation}
\label{s1} \Psi=\sqrt{x} Z_{1/6} \left( {a x^{3}\over
3}\right)\,\sim\, x^{2/3}.
\end{equation}
We note that, in this case, the Bessel function is
$Z_{\nu}=J_{\nu}$ since we must have $|\Psi|\sim 0$ for
$x\rightarrow 0$. This condition would not be satisfied for
$Z_{\nu}=Y_{\nu},H_{\nu}^{(1)}$, because
$Z_{\nu}\rightarrow\infty$ for $x\rightarrow 0$.

In the opposite case $|x|\gg 1$,  Eq.(\ref{wdw1}) becomes
\begin{equation}
\label{wdw3} \Psi''+ \left[
\frac{a^{2}x^2}{(\alpha^2-1)}\right] \Psi =0\,,
\end{equation}
with the solution
\begin{equation}
\label{s3} \Psi(x) = \sqrt{x} Z_{1/4} \left(\frac{a x^2
}{2\sqrt{\alpha^2-1}}\right)\,,
\end{equation}
for $\alpha>1$. In the limit $|x| \rightarrow +\infty$, we get
\begin{equation}
\Psi(x)\approx\left[\frac{2(\alpha^2-1)^{1/4}}{ (\pi a
x)^{1/2}}\right] \exp\left[\pm i\left(\frac{a
x^2}{2\sqrt{\alpha^2-1}}- \frac{3\pi}{4}\right)\right]\,.
\end{equation}
which is clearly an oscillating solution. For $\alpha<1$, the
asymptotic solution is an instanton of the form
\begin{equation}
\Psi(x)\approx\left[\frac{2(|\alpha^2-1|)^{1/4}}{ (\pi
a x)^{1/2}}\right] \exp\left[\pm
\left(\frac{a x^2}{2\sqrt{|\alpha^2-1|}}-
\frac{3\pi}{4}\right)\right]\,,
\end{equation}
which rapidly diverges or converges to zero.

For $|x|\approx 1$, we find
\begin{equation}
\label{wdw4} \Psi'' +2(x-1)\Psi=0\,,
\end{equation}
 and
the solution, in this case, is
\begin{equation}
\label{s5}
 \Psi(x)\approx \sqrt{x-1}
Z_{1/3}\left(\frac{3x^{3/2}}{\sqrt{2}}\right)\,,
\end{equation}
and converges to zero for $x\sim 1$.

Finally, the case $\alpha=1$ is best discussed starting from
Eq.(\ref{wdw1}) directly. We find
\begin{equation}
\label{wdw5}
\Psi''+ a^2\frac{(1-x^2)x^4}{1-2x^2}\Psi=0\,,
\end{equation}
which reduces to
\begin{equation}
\label{wdw6}
\Psi''+a^2x^4\Psi=0\,,\qquad \mbox{for}\;\;x\ll 1\,,
\end{equation}
\begin{equation}
\label{wdw7}
\Psi''+\frac{1}{2}a^2x^4\Psi=0\,,\qquad \mbox{for}\;\;x\gg 1\,,
\end{equation}
\begin{equation}
\label{wdw8}
\Psi''+2a^2(x-1)\Psi=0\,,\qquad \mbox{for}\;\;x\simeq 1\,.
\end{equation}
Eqs.(\ref{wdw6}) and (\ref{wdw7}) are of a well--known type with
solutions
\begin{equation}
\Psi(x)=\sqrt{x}Z_{1/6}\left(\frac{a}{3}x^3\right),\quad
\mbox{and}\quad
\Psi(x)=\sqrt{x}Z_{1/6}\left(\frac{a}{3\sqrt{2}}x^3\right)\;,
\end{equation}
respectively. Eq.(\ref{wdw8}) requires some standard transformations and has
the solution
\begin{equation}
\Psi(x)=\left[2a^2(x-1)\right]^{3/2}Z_{-3/4}
\left(\frac{8}{3}a(x-1)^{3/2}\right)\,.
\end{equation}

\section{\normalsize\bf Discussion}

By looking at the above solutions, we immediately realize that we
have oscillatory behaviours if we are well outside the
superpotential barrier and instantonic solutions if we are near
the singularities (i.e. the zeros) of the superpotential. This
fact can be interpreted as the well known scheme of nucleation
``from nothing" where quantum bubbles are produced in a Lorentzian
region. In our case, deSitter bubbles are produced by breaking
Weyl's conformal symmetry.

If $R(t)$ is interpreted as the scale factor of a bubbly universe,
then $\Psi(R)$ may be connected, in quantum cosmology, to the
probability to obtain specific classes of cosmological models,
which, in our case, are de Sitter universes, as requested by the
inflationary paradigm. However, for $\Psi\rightarrow 0$, this
information is lost. We must, however keep in mind that we are not
in a ``full" quantum gravity regime, where one expects $\Psi$ {\it
to be} exactly a probability amplitude. Here, we can only say that
$\Psi$ is {\it related to} the probability amplitude since the
quantization scheme adopted, as all schemes developed in quantum
gravity up to now, do not yield a Hilbert space.

We also see that the size of the bubble, i.e. its being larger or
smaller than $R_{eq}$, determines its survival and, then, the
probability to give rise to a classical universe. It is important
to remember that $R_{eq}$ depends on the inverse value of an
effective cosmological constant and  is therefore related to the
energy of the vacuum from which the bubbles are produced. In other
words, the effective cosmological constant determines the
dynamics. Moreover, the oscillatory behaviours of the wave
function well fit Vilenkin's {\it no boundary conditions} to
obtain classical trajectories in the minisuperspace
\cite{vilenkin}.  An oscillating wavefunction is capable of {\it
selecting} first integrals of motion from which classical
cosmological models are derived, in the semiclassical
approximation of canonical quantum gravity \cite{capozziello}. In
the present case, the quantum production of de Sitter bubbles
gives a non--null probability to obtain initial conditions for
inflation. In summary, the breaking of  Weyl symmetry provides a
standard of length in Riemannian geometry and, in addition, the
initial conditions favourable to the production of observable
universes.

\vspace{3. mm}

\begin{center}
{\bf Aknowledgements}
\end{center}
Research supported by MURST ex 40\% and 60\% art. 65 D.P.R.
382/80. G.L. thanks the financial support of EU (P.O.M.1994/1999).

\end{document}